\documentclass[sigconf]{acmart}
\AtBeginDocument{%
  \providecommand\BibTeX{{%
    \normalfont B\kern-0.5em{\scshape i\kern-0.25em b}\kern-0.8em\TeX}}}

\setcopyright{acmcopyright}
\copyrightyear{2023}
\acmYear{2023}
\acmDOI{XXXXXXX.XXXXXXX}

\acmConference[ISAV '23]{In Situ Infrastructures for Enabling Extreme-scale Analysis and Visualization}{November 13,
  2023}{Denver, CO}
\acmPrice{15.00}
\acmISBN{978-1-4503-XXXX-X/18/06}

\begin{document}

\title{State of In Situ Visualization in Simulations:\\We are fast. But are we inspiring?}

\author{Axel Huebl}
\email{{axelhuebl,jlvay}@lbl.gov}
\orcid{0000-0003-1943-7141}
\author{Arianna Formenti}
\orcid{0000-0002-7887-9313}
\author{Marco Garten}
\orcid{0000-0001-6994-2475}
\author{Jean-Luc Vay}
\orcid{0000-0002-0040-799X}
\affiliation{%
  \institution{Lawrence Berkeley National Laboratory}
  \streetaddress{1 Cyclotron Rd}
  \city{Berkeley}
  \state{California}
  \country{USA}
  \postcode{94720}
}


\begin{abstract}
  Visualization of dynamic processes in scientific high-performance computing is an immensely data intensive endeavor.
  Application codes have recently demonstrated scaling to full-size Exascale machines, and generating high-quality data for visualization is consequently on the machine-scale, easily spanning 100s of TBytes of input to generate a single video frame.
  In situ visualization, the technique to consume the many-node decomposed data in-memory, as exposed by applications, is the dominant workflow.
  Although in situ visualization has achieved tremendous progress in the last decade, scaling to system-size together with the application codes that produce its data, there is one important question that we cannot skip: is what we produce insightful and inspiring?
\end{abstract}



\begin{teaserfigure}
  \includegraphics[width=\textwidth]{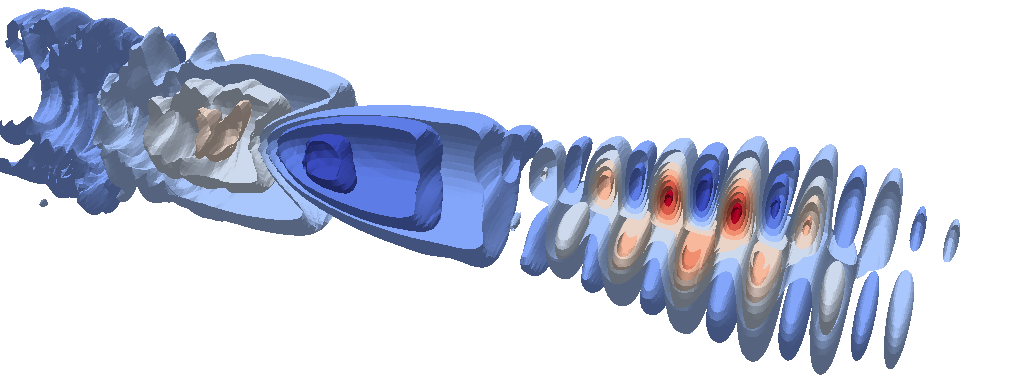}
  \caption{WarpX in situ visualization with Ascent and VTK-m.
  The figure shows a laser-wakefield accelerator stage (left), driven by a laser pulse (right) that travels to the bottom right.
  Iso-contours show the longitudinal electric field used in particle acceleration.
  The scenario is modeled and visualized in a Lorentz-boosted reference frame \cite{VayPRL2007,VayJCP2011,VayPoP2011}.
  If this was not \textit{in situ}, the authors would be able to transform to a different reference frame, add multiple light sources, cast hard and soft shadows, select some iso-contours for semi-transparent representation, and would add more triangles to smooth the generated iso-contours.}
  \label{fig:teaser}
\end{teaserfigure}

\received{11 August 2023}
\received[revised]{30 September 2023}

\maketitle

\section{Introduction}

In situ visualization is a tremendously powerful workflow to generate insight into the largest simulations run today.
Recently, the 2022 Gordon Bell Prize-winning application WarpX~\cite{FedeliHuebl2022} was used to run in situ visualization on 552 nodes of the Frontier supercomputer~\cite{MarsagliaKPP32023}.

Immediate visualization of simulation dynamics at scale, from various camera angles, is powerful and helpful, providing answers to domain-science questions such as:
Is a simulation evolving as planned?
Are numerical options and resolution sufficiently set?
Are any hardware or software issues/bugs appearing at scale?
Yet, the scientifically most important question is:
Does the visualization develop insight?

Gaining scientific insight from simulations is a complex and iterative process, with domain scientists connecting existing theory, empirical evidence and data from experiments and simulations.
Visualizations can produce qualitative and quantitative representations of the dynamics at play.
These representations can solidify understanding, guide the theoretical model building, help testing approximations and assumptions.
An attractive visualization does help to communicate results and might inspire new scientific ideas.

Particularly for the latter part, domain scientists and audiences will compare the quality of their visualization with the state-of-the-art seen in everyday life: movies, games, advertising, etc.
That is a high bar, given photo-realistic capabilities in these industries at high frame rates.
Based on these expectations, can we produce in situ visualizations of scientific data that can be awe-inspiring and stimulate our minds?
And - how much costs and/or scalability penalty are we willing to trade for this in high-performance computing?

\section{Scalable Methods Wanted}

Many algorithms offered in contemporary visualization frameworks~\cite{ascent,Larsen2022,vtkm,Meyer2023} are able to exploit some locality, e.g., by domain decomposing ray traces and iso-contour searches, composing results later on~\cite{icet}.
Yet, advanced visualization techniques for casting shadows, tracing reflections, sorting collisions with objects, etc. are notoriously non-local and are thus challenging for multi-GPU implementations.
Even volume-rendering more than one spatially overlapping source is non-trivial to do \textit{in situ}, since established methods depend on a sampling technique that is hard to scale~\cite{MarsagliaPrivComm}.
Additionally, many visualization techniques that scientists can use on single-node implementations would be highly desirable as distributed implementations for in situ frameworks:
Taking Figure~\ref{fig:teaser} as an example, if this was not \textit{in situ} generated, the authors would add multiple light sources, cast hard and soft shadows, select some iso-contours for semi-transparent representation, and would smooth the generated iso-contours, by adding additional triangles that interpolate beyond the original resolution of the data source.

Consequently, there is a continued need for new, innovative, scalable in situ visualization methods.
Both fast, low-overhead and higher-overhead (yet scalable), high-quality methods are needed.
With respect to scalability, maybe there are tricks one can lend from other communities to generate artificial locality:
occlude far-from-focus parts with mist as in gaming, simplify shadow masks and reflections, or aggressively exploit the adaptive resolution of mesh-refined data sources.
Additionally, successful in situ implementations and workflows can likely be enhanced and benefit from evolution through standardization of APIs, vendor abstractions, render scene control and data descriptions, e.g., \cite{Conduit,ANARI,openUSD}.

\section{Selected In Situ Visualization Needs}

Adding to the challenges of addressing expectations set from offline rendering for in situ visualization, we surveyed the Beam, Plasma \& Accelerator Simulation Toolkit (BLAST)~\cite{blast,FedeliHuebl2022,Diederichs2022,ImpactX} codes and identified three selected needs specific to in situ visualization.

First, we noticed that domain scientists have to relearn how to express rendering scene descriptions for each in situ tool.
Standardization is needed \cite{openUSD}.
Another approach might be domain-specific options in the simulation input language, automating the creation of visualization-configuration templates with mostly defaulted options - ready to be configured further for details by the inclined scientists when needed.

Second, video generation of iso-contours, glyphs (e.g., vectors placed in space), etc. often create ``flicker'' effects for surfaces and pointing of objects, simply based on the roughness of simulation data and steps selected for visualization.
Research into transitions (or animations) between key/data frames with low memory overhead for HPC could be beneficial to reduce such effects.

Third, we also identified a commonly used algorithmic and simulation pattern for which in situ visualization would be ideally suited, but are not aware of any implemented solution yet:
rendering of spatially-sliced data pipelines.
In a large class of modeling codes, efficient solutions can be calculated by splitting the 3D domain over one axis.
Instead of advancing the whole domain by an update, algorithms update a slice of the domain, e.g., from the back to the front of the 3D domain, and parallelize for the third spatial axis in \textit{time}.
Without spatially sliced rendering tools, a large number of algorithms and codes currently need to fall back to costly data output to ``reconstruct'' the spatial data domain that is required at once in offline visualization.
Examples in laser-plasma and accelerator physics are the boosted frame technique \cite{VayPRL2007,VayJCP2011,VayPoP2011} as shown in figure~\ref{fig:teaser} (a more meaningful representation would transform slice-wise to the laboratory frame), the quasi-static method \cite{Diederichs2022}, or representations in reference trajectory space instead of time and space \cite{ImpactX}.
\newline

We believe addressing these challenges is timely and resulting in situ visualization will provide insight and inspiration for scientists.

\begin{acks}
This research was supported by the Exascale Computing Project (17-SC-20-SC), a collaborative effort of the U.S. Department of Energy Office of Science and the National Nuclear Security Administration. 
This research was supported by the U.S. Department of Energy, Office of Science, Office of Advanced Scientific Computing Research and Office of High Energy Physics, Scientific Discovery through Advanced Computing (SciDAC) program.
This research used resources of the Oak Ridge Leadership Computing Facility at the Oak Ridge National Laboratory, which is supported by the Office of Science of the U.S. Department of Energy under Contract No. DE-AC05-00OR22725.
\end{acks}

\bibliographystyle{ACM-Reference-Format}
\bibliography{references}


\begin{thebibliography}{17}


\ifx \showCODEN    \undefined \def \showCODEN     #1{\unskip}     \fi
\ifx \showDOI      \undefined \def \showDOI       #1{#1}\fi
\ifx \showISBNx    \undefined \def \showISBNx     #1{\unskip}     \fi
\ifx \showISBNxiii \undefined \def \showISBNxiii  #1{\unskip}     \fi
\ifx \showISSN     \undefined \def \showISSN      #1{\unskip}     \fi
\ifx \showLCCN     \undefined \def \showLCCN      #1{\unskip}     \fi
\ifx \shownote     \undefined \def \shownote      #1{#1}          \fi
\ifx \showarticletitle \undefined \def \showarticletitle #1{#1}   \fi
\ifx \showURL      \undefined \def \showURL       {\relax}        \fi
\providecommand\bibfield[2]{#2}
\providecommand\bibinfo[2]{#2}
\providecommand\natexlab[1]{#1}
\providecommand\showeprint[2][]{arXiv:#2}

\bibitem[Diederichs et~al\mbox{.}(2022)]%
        {Diederichs2022}
\bibfield{author}{\bibinfo{person}{Severin Diederichs}, \bibinfo{person}{Carlo
  Benedetti}, \bibinfo{person}{Axel Huebl}, \bibinfo{person}{Rremi Lehe},
  \bibinfo{person}{Andrew Myers}, \bibinfo{person}{Alexander Sinn},
  \bibinfo{person}{Jean-Luc Vay}, \bibinfo{person}{Weiqun Zhang}, {and}
  \bibinfo{person}{Maxence Thévenet}.} \bibinfo{year}{2022}\natexlab{}.
\newblock \showarticletitle{HiPACE++: A portable, 3D quasi-static
  particle-in-cell code}.
\newblock \bibinfo{journal}{\emph{Computer Physics Communications}}
  \bibinfo{volume}{278} (\bibinfo{year}{2022}), \bibinfo{pages}{108421}.
\newblock
\showISSN{0010-4655}
\urldef\tempurl%
\url{https://doi.org/10.1016/j.cpc.2022.108421}
\showDOI{\tempurl}


\bibitem[Fedeli et~al\mbox{.}(2022)]%
        {FedeliHuebl2022}
\bibfield{author}{\bibinfo{person}{Luca Fedeli}, \bibinfo{person}{Axel Huebl},
  \bibinfo{person}{France Boillod-Cerneux}, \bibinfo{person}{Thomas Clark},
  \bibinfo{person}{Kevin Gott}, \bibinfo{person}{Conrad Hillairet},
  \bibinfo{person}{Stephan Jaure}, \bibinfo{person}{Adrien Leblanc},
  \bibinfo{person}{Rémi Lehe}, \bibinfo{person}{Andrew Myers},
  \bibinfo{person}{Christelle Piechurski}, \bibinfo{person}{Mitsuhisa Sato},
  \bibinfo{person}{Neïl Zaim}, \bibinfo{person}{Weiqun Zhang},
  \bibinfo{person}{Jean-Luc Vay}, {and} \bibinfo{person}{Henri Vincenti}.}
  \bibinfo{year}{2022}\natexlab{}.
\newblock \showarticletitle{Pushing the Frontier in the Design of Laser-Based
  Electron Accelerators with Groundbreaking Mesh-Refined Particle-In-Cell
  Simulations on Exascale-Class Supercomputers}. In
  \bibinfo{booktitle}{\emph{SC22: International Conference for High Performance
  Computing, Networking, Storage and Analysis}}. \bibinfo{publisher}{IEEE},
  \bibinfo{address}{Dallas (TX), USA}, \bibinfo{pages}{1--12}.
\newblock
\urldef\tempurl%
\url{https://doi.org/10.1109/SC41404.2022.00008}
\showDOI{\tempurl}


\bibitem[Harrison et~al\mbox{.}(2022)]%
        {Conduit}
\bibfield{author}{\bibinfo{person}{Cyrus Harrison}, \bibinfo{person}{Matthew
  Larsen}, \bibinfo{person}{Brian~S. Ryujin}, \bibinfo{person}{Adam Kunen},
  \bibinfo{person}{Arlie Capps}, {and} \bibinfo{person}{Justin Privitera}.}
  \bibinfo{year}{2022}\natexlab{}.
\newblock \showarticletitle{Conduit: A Successful Strategy for Describing and
  Sharing Data In Situ}. In \bibinfo{booktitle}{\emph{2022 IEEE/ACM
  International Workshop on In Situ Infrastructures for Enabling Extreme-Scale
  Analysis and Visualization (ISAV)}}. \bibinfo{publisher}{IEEE},
  \bibinfo{address}{Dallas (TX), USA}, \bibinfo{pages}{1--6}.
\newblock
\urldef\tempurl%
\url{https://doi.org/10.1109/ISAV56555.2022.00006}
\showDOI{\tempurl}


\bibitem[Huebl et~al\mbox{.}(2022)]%
        {ImpactX}
\bibfield{author}{\bibinfo{person}{Axel Huebl}, \bibinfo{person}{Remi Lehe},
  \bibinfo{person}{Chad Mitchell}, \bibinfo{person}{Ji Qiang},
  \bibinfo{person}{Robert Ryne}, \bibinfo{person}{Ryan Sandberg}, {and}
  \bibinfo{person}{Jean-Luc Vay}.} \bibinfo{year}{2022}\natexlab{}.
\newblock \showarticletitle{Next Generation Computational Tools for the
  Modeling and Design of Particle Accelerators at Exascale}.
\newblock \bibinfo{journal}{\emph{Proceedings of the 5th North American
  Particle Accelerator Conference}}  \bibinfo{volume}{NAPAC2022}
  (\bibinfo{year}{2022}), \bibinfo{pages}{USA}.
\newblock
\urldef\tempurl%
\url{https://doi.org/10.18429/JACOW-NAPAC2022-TUYE2}
\showDOI{\tempurl}


\bibitem[Larsen et~al\mbox{.}(2017)]%
        {ascent}
\bibfield{author}{\bibinfo{person}{Matthew Larsen}, \bibinfo{person}{James
  Ahrens}, \bibinfo{person}{Utkarsh Ayachit}, \bibinfo{person}{Eric Brugger},
  \bibinfo{person}{Hank Childs}, \bibinfo{person}{Berk Geveci}, {and}
  \bibinfo{person}{Cyrus Harrison}.} \bibinfo{year}{2017}\natexlab{}.
\newblock \showarticletitle{The ALPINE In Situ Infrastructure: Ascending from
  the Ashes of Strawman}. In \bibinfo{booktitle}{\emph{Proceedings of the In
  Situ Infrastructures on Enabling Extreme-Scale Analysis and Visualization}}
  (Denver, CO, USA) \emph{(\bibinfo{series}{ISAV'17})}.
  \bibinfo{publisher}{Association for Computing Machinery},
  \bibinfo{address}{New York, NY, USA}, \bibinfo{pages}{42–46}.
\newblock
\showISBNx{9781450351393}
\urldef\tempurl%
\url{https://doi.org/10.1145/3144769.3144778}
\showDOI{\tempurl}


\bibitem[Larsen et~al\mbox{.}(2022)]%
        {Larsen2022}
\bibfield{author}{\bibinfo{person}{Matthew Larsen}, \bibinfo{person}{Eric
  Brugger}, \bibinfo{person}{Hank Childs}, {and} \bibinfo{person}{Cyrus
  Harrison}.} \bibinfo{year}{2022}\natexlab{}.
\newblock \showarticletitle{{Ascent: A Flyweight In Situ Library for Exascale
  Simulations}}.
\newblock In \bibinfo{booktitle}{\emph{{In Situ Visualization For Computational
  Science}}}. \bibinfo{publisher}{Mathematics and Visualization book series
  from Springer Publishing}, \bibinfo{address}{Cham, Switzerland},
  \bibinfo{pages}{255 -- 279}.
\newblock


\bibitem[{LBNL} et~al\mbox{.}(2023)]%
        {blast}
\bibfield{author}{\bibinfo{person}{{LBNL}}, \bibinfo{person}{{LLNL}},
  \bibinfo{person}{{SLAC}}, \bibinfo{person}{{CEA-LIDYL}},
  \bibinfo{person}{{DESY}}, \bibinfo{person}{{U Hamburg}},
  \bibinfo{person}{{CERN}}, \bibinfo{person}{{TAE}}, {et~al\mbox{.}}}
  \bibinfo{year}{2023}\natexlab{}.
\newblock \bibinfo{title}{{Beam, Plasma \& Accelerator Simulation Toolkit
  (BLAST)}}.
\newblock
\newblock
\urldef\tempurl%
\url{https://blast.lbl.gov}
\showURL{%
\tempurl}


\bibitem[Marsaglia(2023)]%
        {MarsagliaPrivComm}
\bibfield{author}{\bibinfo{person}{Nicole Marsaglia}.}
  \bibinfo{year}{2023}\natexlab{}.
\newblock \bibinfo{howpublished}{personal communication}.
\newblock


\bibitem[Marsaglia et~al\mbox{.}(2023)]%
        {MarsagliaKPP32023}
\bibfield{author}{\bibinfo{person}{Nicole Marsaglia}, \bibinfo{person}{Cyrus
  Harrison}, \bibinfo{person}{Matthew Larsen}, \bibinfo{person}{Axel Huebl},
  {and} \bibinfo{person}{Jean-Luc Vay}.} \bibinfo{year}{2023}\natexlab{}.
\newblock \bibinfo{booktitle}{\emph{KPP-3 Artifact for WarpX <> ALPINE: Ascent
  + VTK-m}}.
\newblock DOE Exascale Computing Project.
\newblock
\urldef\tempurl%
\url{https://doi.org/10.5281/zenodo.8226853}
\showDOI{\tempurl}


\bibitem[Meyer et~al\mbox{.}(2023)]%
        {Meyer2023}
\bibfield{author}{\bibinfo{person}{Felix Meyer}, \bibinfo{person}{Benjamin
  Hernandez}, \bibinfo{person}{Richard Pausch}, \bibinfo{person}{Ren\'{e}
  Widera}, \bibinfo{person}{David Gro\ss{}}, \bibinfo{person}{Sergei
  Bastrakov}, \bibinfo{person}{Axel Huebl}, \bibinfo{person}{Guido Juckeland},
  \bibinfo{person}{Jeffrey Kelling}, \bibinfo{person}{Matt Leinhauser},
  \bibinfo{person}{David Rogers}, \bibinfo{person}{Ulrich Schramm},
  \bibinfo{person}{Klaus Steiniger}, \bibinfo{person}{Stefan Gumhold},
  \bibinfo{person}{Jeff Young}, \bibinfo{person}{Michael Bussmann},
  \bibinfo{person}{Sunita Chandrasekaran}, {and} \bibinfo{person}{Alexander
  Debus}.} \bibinfo{year}{2023}\natexlab{}.
\newblock \showarticletitle{Hardware-Agnostic Interactive Exascale In Situ
  Visualization of Particle-In-Cell Simulations}. In
  \bibinfo{booktitle}{\emph{Proceedings of the Platform for Advanced Scientific
  Computing Conference}} (Davos, Switzerland) \emph{(\bibinfo{series}{PASC
  '23})}. \bibinfo{publisher}{Association for Computing Machinery},
  \bibinfo{address}{New York, NY, USA}, Article \bibinfo{articleno}{9},
  \bibinfo{numpages}{14}~pages.
\newblock
\showISBNx{9798400701900}
\urldef\tempurl%
\url{https://doi.org/10.1145/3592979.3593408}
\showDOI{\tempurl}


\bibitem[Moreland(2009)]%
        {icet}
\bibfield{author}{\bibinfo{person}{Kenneth Moreland}.}
  \bibinfo{year}{2009}\natexlab{}.
\newblock \bibinfo{title}{IceT Users’ Guide and Reference}.
\newblock \bibinfo{howpublished}{Sandia Report SAND2009-3170}.
\newblock
\urldef\tempurl%
\url{https://www.osti.gov/servlets/purl/970256}
\showURL{%
\tempurl}


\bibitem[Moreland et~al\mbox{.}(2016)]%
        {vtkm}
\bibfield{author}{\bibinfo{person}{Kenneth Moreland},
  \bibinfo{person}{Christopher Sewell}, \bibinfo{person}{William Usher},
  \bibinfo{person}{Li-ta Lo}, \bibinfo{person}{Jeremy Meredith},
  \bibinfo{person}{David Pugmire}, \bibinfo{person}{James Kress},
  \bibinfo{person}{Hendrik Schroots}, \bibinfo{person}{Kwan-Liu Ma},
  \bibinfo{person}{Hank Childs}, \bibinfo{person}{Matthew Larsen},
  \bibinfo{person}{Chun-Ming Chen}, \bibinfo{person}{Robert Maynard}, {and}
  \bibinfo{person}{Berk Geveci}.} \bibinfo{year}{2016}\natexlab{}.
\newblock \showarticletitle{VTK-m: Accelerating the Visualization Toolkit for
  Massively Threaded Architectures}.
\newblock \bibinfo{journal}{\emph{IEEE Computer Graphics and Applications}}
  \bibinfo{volume}{36}, \bibinfo{number}{3} (\bibinfo{year}{2016}),
  \bibinfo{pages}{48--58}.
\newblock
\urldef\tempurl%
\url{https://doi.org/10.1109/MCG.2016.48}
\showDOI{\tempurl}


\bibitem[Pixar(2016)]%
        {openUSD}
\bibfield{author}{\bibinfo{person}{Pixar}.} \bibinfo{year}{2016}\natexlab{}.
\newblock \bibinfo{title}{Universal Scene Description}.
\newblock
\newblock
\urldef\tempurl%
\url{https://openusd.org}
\showURL{%
\tempurl}


\bibitem[Stone et~al\mbox{.}(2022)]%
        {ANARI}
\bibfield{author}{\bibinfo{person}{John~E. Stone}, \bibinfo{person}{Kevin~S.
  Griffin}, \bibinfo{person}{Jefferson Amstutz}, \bibinfo{person}{David~E.
  DeMarle}, \bibinfo{person}{William~R. Sherman}, {and}
  \bibinfo{person}{Johannes Günther}.} \bibinfo{year}{2022}\natexlab{}.
\newblock \showarticletitle{ANARI: A 3-D Rendering API Standard}.
\newblock \bibinfo{journal}{\emph{Computing in Science \& Engineering}}
  \bibinfo{volume}{24}, \bibinfo{number}{2} (\bibinfo{year}{2022}),
  \bibinfo{pages}{7--18}.
\newblock
\urldef\tempurl%
\url{https://doi.org/10.1109/MCSE.2022.3163151}
\showDOI{\tempurl}


\bibitem[Vay(2007)]%
        {VayPRL2007}
\bibfield{author}{\bibinfo{person}{Jean-Luc Vay}.}
  \bibinfo{year}{2007}\natexlab{}.
\newblock \showarticletitle{Noninvariance of Space- and Time-Scale Ranges under
  a Lorentz Transformation and the Implications for the Study of Relativistic
  Interactions}.
\newblock \bibinfo{journal}{\emph{Phys. Rev. Lett.}}  \bibinfo{volume}{98}
  (\bibinfo{date}{Mar} \bibinfo{year}{2007}), \bibinfo{pages}{130405}.
\newblock
Issue 13.
\urldef\tempurl%
\url{https://doi.org/10.1103/PhysRevLett.98.130405}
\showDOI{\tempurl}


\bibitem[Vay et~al\mbox{.}(2011a)]%
        {VayJCP2011}
\bibfield{author}{\bibinfo{person}{Jean-Luc Vay}, \bibinfo{person}{Cameron~G.R.
  Geddes}, \bibinfo{person}{Estelle Cormier-Michel}, {and}
  \bibinfo{person}{David~P. Grote}.} \bibinfo{year}{2011}\natexlab{a}.
\newblock \showarticletitle{Numerical methods for instability mitigation in the
  modeling of laser wakefield accelerators in a Lorentz-boosted frame}.
\newblock \bibinfo{journal}{\emph{J. Comput. Phys.}} \bibinfo{volume}{230},
  \bibinfo{number}{15} (\bibinfo{year}{2011}), \bibinfo{pages}{5908--5929}.
\newblock
\showISSN{0021-9991}
\urldef\tempurl%
\url{https://doi.org/10.1016/j.jcp.2011.04.003}
\showDOI{\tempurl}


\bibitem[Vay et~al\mbox{.}(2011b)]%
        {VayPoP2011}
\bibfield{author}{\bibinfo{person}{Jean-Luc Vay}, \bibinfo{person}{Cameron
  G.~R. Geddes}, \bibinfo{person}{Estelle Cormier-Michel}, {and}
  \bibinfo{person}{David~P. Grote}.} \bibinfo{year}{2011}\natexlab{b}.
\newblock \showarticletitle{{Effects of hyperbolic rotation in Minkowski space
  on the modeling of plasma accelerators in a Lorentz boosted frame}}.
\newblock \bibinfo{journal}{\emph{Physics of Plasmas}} \bibinfo{volume}{18},
  \bibinfo{number}{3} (\bibinfo{date}{03} \bibinfo{year}{2011}),
  \bibinfo{pages}{030701}.
\newblock
\showISSN{1070-664X}
\urldef\tempurl%
\url{https://doi.org/10.1063/1.3559483}
\showDOI{\tempurl}


\end{thebibliography}










\end{document}